# Comparative study of Joint Image Encryption and Compression Schemes: A Review

Behrooz Khadem, Reza Ahmadian

*Abstract*—With the development of imaging methods in wireless communications, enhancing the security and efficiency of image transfer requires image compression and encryption schemes. In conventional methods, encryption and compression are two separate processes, therefore an adversary can organize his attack more simply but if these two processes are combined, the output uncertainty increases. As a result, adversaries face more difficulties, and schemes will be more secure. This paper introduces a number of the most important criteria for the efficiency and security evaluation of joint image encryption and compression (JIEC) schemes. These criteria were then employed to compare the schemes. The comparison results were analysed to propose suggestions and strategies for future research to develop secure and efficient JIEC schemes.

## I. INTRODUCTION

Nowadays, the explosion of image transfer requests in wireless communications for medical and paramedical [1], environmental protection and weather forecast [2], agriculture [3], military and security [4], supervision and inspection in various industries [5], architecture, and transportation systems [5] and impassable mines and sites [6] purposes has caused much more attention to image compression and encryption. On the one hand, the studies of image compression have focused on image size reduction for easy storage and quick transfer. On the other hand, many studies have been conducted on encryption schemes and systems with acceptable diffusion and confusion features to ensure image confidentiality of users [7]. As compression methods are incurious to the security of transferred images, encryption schemes obviously put no emphasis on compression due to focusing on image confidentiality. Thus, there has been growing necessity for JIEC schemes in recent years. In conventional encryption and compression methods, the two processes are performed totally distinctively and independently. For instance, a block cipher such as AES is applied to JPEG images. In this case, an adversary can focus perfectly on the encryption output without considering compression and organize an attack due to the occurrence of scheme incompatibility [8]. On the contrary, if these two processes are combined, the cipher output image will be more complicated so that adversaries will face more difficulties. Hence, JIEC schemes have received much more attention and research emphasis.

In addition to certain requirements such as online connectivity, cipher image uniformity, and compression for JIEC schemes, there are other important prerequisites such as protocol constraints as well as physical constraints including dimensions, weight, frequency, power, range, and antenna. The communication channel bandwidth and information transfer rate are also important. All of these factors have significant roles in selecting and combining compression and encryption to transfer images as fast as possible in addition to maintaining security and ensuring the best quality of image recovery. Sometimes, the use of a highly secure encryption method may cause incompatibility if compressibility increases. On the contrary, an adversary can sometimes perform decryption more easily due to the existing redundancies when compression is weak. It is usually difficult to resolve these challenges and incompatibilities in different applications of wireless communications systems.

### A. Necessity and Importance of JIEC schemes

Wireless communications are now widely used for imaging and image transfer in different areas. In all of these applications, it is necessary to send high-quality, low-size, and (in some cases) secure images. Unlike text messages, images are characterized by specific features such as high size, redundancies and high correlation of pixels making the execution of conventional encryption methods very slow. In addition, image applications need specific requirements such as online connectivity, image uniformity, compression, no data lost, and proper bandwidth for transfer [9, 10].

Regarding the above-mentioned requirements, many studies have been conducted to develop JIEC schemes. This paper compared some of the most important and latest JIEC schemes (2013-2017) to enable researchers to select their favorable schemes. Some suggestions were also made in this paper to help researchers develop new, secure, and efficient JIEC schemes. Table 1 shows a number of abbreviations used in this paper.

The structure of this paper is as follows. Section 2 reviews a number of the latest JIEC schemes. Section 3 introduces some of the most common and important evaluation criteria for image encryption and compression. In section 4, a comparison is drawn between the schemes introduced in section 2 based on the criteria discussed in section 3 to analyze the advantages and disadvantages of the schemes. According to the results in section 4, some recommendations are provided for the production or development of secure and efficient JIEC schemes in section 5.

## II. LITERATURE REVIEW

In this section, some of the well-known and latest JIEC schemes are introduced.

Zhu et al. (2013) proposed a JIEC scheme [11]. They first proposed a 2D hyper-chaos discrete nonlinear dynamic system to scramble the sent image. Then they employed the Chinese reminder theorem (CRT) to diffuse and compress the scrambled image. Their scheme could be used for changing

Behrooz Khadem, Assistant Professor, Imam Hossein Comprehensive University, (e-mail: Bkhadem@ihu.ac.ir).
Reza Ahmadian, MSc Student of Encryption and Secure Telecommunications, Imam Hossein Comprehensive University, (e-mail: Rahmadian@ihu.ac.ir).

the plain image information and compressing it at any given CR such as k.

Table 1. Some used abbreviations

| Metric | Definition |
|---|---|
| IWT | Integer Wavelet Transform |
| DWT | Discrete Wavelet Transform |
| DCT | Discrete Cosine Transform |
| DFT | Fourier Wavelet Transform |
| CRT | Chinese Remainder Transform |
| CS | Compressive Sensing |
| PSNR | Peak Signal-to-Noise Ratio |
| SPIHT | Set Partitioning in Hierarchical Trees |
| NPCR | Number of Changing Pixel Rate |
| UACI | Unified Average Changed Intensity |
| MAE | Mean Absolute Error |
| CR | Compression Rate |
| bpp | bit per pixel |
| MSE | Mean Squared Error |
| JIEC | Joint Image Encryption and Compression |

Zhu et al. (2016) introduced a JIEC scheme based on 2D compressive sensing and Chen's hyper-chaos system [12]. They first employed the compressive sensing technique for compression. Then they benefited from a large made-up Hadamard matrix for JIEC. In the next step, the compressed image was re-encrypted with the hyper-chaotic system, in which pixel rotation was used effectively.

Chen et al. (2016) proposed a JIEC scheme consisting of a Kronecker compressive sensing structure and CA [13]. They first used CA to scramble the plain image. In their scheme, in order to create calculation complexity and reduce memory usage, they employed the Kronecker compressive sensing method for compression. They also used DFT, DWT, and DCT separately to generate compression coefficients.

Tang et al. (2016) proposed a new JIEC scheme [14]. The scheme employed the DCT dictionary to sparsely represent the color image and then combined it with the encryption algorithm based on the hyper-chaotic system to achieve image compression and encryption simultaneously. The image is compressed by folding the color image using the nature of the orthogonal projection matrix and the relationship among image data, DCT dictionary and DCT. The diffusion and scrambling image encryption algorithm based on the hyper-chaotic system is designed to encrypt the image. In the process of the folded compression, the image is encrypted from three aspects: Firstly, a scrambling method for the position of effective coefficients is proposed to reach the goal of eliminating the correlation between pixels in one block. Then in order to encrypt the coefficients, the sparse coefficients are controlled by the hyper-chaotic system, and finally, a diffusion method for the folded image is proposed to encrypt the final data.

Hamdi et al. (2017) introduced a considerably secure and high-speed join encryption and compression scheme [15]. In their scheme, encryption and compression were based on the wavelet transform, Chirikov standard map, and SPIHT. They first employed SPIHT to scramble all of the input pixels in order to eliminate pixel correlation. Then unimportant pixels were deleted by using the wavelet transform, and the chaotic map was used for encryption. Finally, they diffused the image bits.

Tang et al. (2017) proposed a new JIEC scheme based on the chaotic map [8]. Their scheme benefited from waste-free SPIHT compression based on IWT. They employed different chaotic maps such as Hénon, Lorenz, and logistic to encrypt wavelet and SPIHT coefficients. They used Lorenz map to confuse wavelet coefficients. They also used the logistic map to encrypt the output SPIHT sequence. Finally, they benefited from Hénon map to scramble the wavelet coefficients. The encrypted wavelet coefficients were recursively used as the input for the next wavelet coefficients.

Zhang et al. (2017) proposed a recursive scheme based on SPIHT, IWT, and a hyper-chaotic system, in which compression was based on IWT and SPIHT [16]. In their scheme, IWT coefficients were generated first. Then the coefficients were encrypted through confusion and diffusion developed by the Rabinovich hyper-chaotic map. In the next step, the cipher image pixels were embedded into SPIHT for recompression. Finally, permutation and diffusion were applied to the output. In addition to the Rabinovich hyper-chaotic map, nonlinear inverted operations were used along with the SHA-256 process to enhance security. Table 2 introduces a brief overview of encryption and compression methods used in different schemes. The next section presents some of the most important criteria for the security and efficiency of JIEC schemes.

Ahmad et al. (2017) proposed a JIEC scheme based on chaotic maps and orthogonal matrices [17]. To obtain a random orthogonal matrix via the Gram Schmidt algorithm, a well-known nonlinear chaotic map was used in to diffuse pixels values of a plaintext image. In the process of block-wise random permutation, the logistic map was employed followed by the diffusion process. They showed by experimental results and security analyses such as key space, differential and statistical attacks that the proposed scheme was secure enough and robust against channel noise and JPEG compression.

Xiaoyong et al. (2017) proposed an efficient and simple encryption and compression scheme based on the algorithm of the generalized knight's-tour, DCT and Chen chaotic maps [18]. In the first step the generalized knight's tour algorithm was utilized to scramble the pixels while the data correlation preserved. Then, the chaotic system was used to generate a pseudorandom permutation to encrypt the part of coefficients from DCT for diffusion. In encoding procedures, they used NGKT matrices generated by Semi Ham algorithm to scramble the plain image and utilize DCT and quantization coding to compress the image with a high compression ratio. Then, diffusion was achieved by part encrypting the DCT coefficients based on Chen chaotic system.

### III. JIEC SCHEMES EVALUATION CRITERIA

Different criteria are employed to compare JIEC schemes. These criteria are generally divided into two classes, security and efficiency. Some of them are addressed in this section.

#### A. Security Criteria

Security criteria include the key space size, cipher image histogram, correlation, entropy, key sensitivity of the cipher image, and resistance to statistical and differential attacks.

The large size of the encryption key space is a necessary prerequisite to JIEC schemes. Empirically, the key space should be greater than $2^{100}$ to resist exhaustive search attacks [17].

In a good JIEC scheme, the cipher image histogram should be relatively uniform. In other words, every gray level has nearly the same occurrence possibility in the cipher image [19].

In JIEC schemes, the correlation of the cipher image pixels should be very low. In other words, the cipher image should include no information about the plain image pattern. This criterion includes horizontal, vertical, and diagonal correlations of image pixels, which are determined separately for the plain and cipher images. Equation 1 is usually used in these schemes to determine and compare the coefficient of correlation. In this equation, A and B indicate the brightness of two adjacent pixels, and m×n shows the number of pixels on the image. Moreover, $\bar{A}$ and $\bar{B}$ refer to the mean brightness of the output image pixels [20].

$$r = \frac{\sum_m \sum_n (A_{mn} - \bar{A})(B_{mn} - \bar{B})}{\sqrt{\left(\sum_m \sum_n (A_{mn} - \bar{A})^2\right)\left(\sum_m \sum_n (B_{mn} - \bar{B})^2\right)}} \quad (1)$$

Equation 2 can be employed to determine the randomness of pixels on the output image with regard to the entropy of their gray level. In Equation 2, the number of pixels range between 0 and 255 on the output image. They should have a relatively equal occurrence probability. In this equation, $p(x_i)$ indicate the occurrence probability of every pixel ($i = 0$ to $i = 255$). The ideal value of entropy is usually assumed 8 and can be determined through Equation 2 [21].

$$H = -\sum_{i=0}^{Q-1} P(x_i) \log_2 \frac{1}{p(x_i)} \quad (2)$$

The output image of a good JIEC scheme should be sensitive to the encryption key so that large changes can be seen on the output image if a few changes are made in the key bits. To determine the sensitivity, the image should be encrypted first with the master key and then with another key (differing one bit from the master key). The two resultant output images are compared in every pixel, and the result is expressed in a percentage [22].

In every JIEC scheme, changing one pixel on the plain image should make large changes on the cipher image to resist differential attacks and the attacks on the selected plain image and the known plain image. In a differential attack, an adversary attempts to make a few changes in the pixels of the input image to see the result in the output image with the purpose of understanding the relationship between input and output images. Usually, UACI and NPCR are employed to determine resistance to differential attacks on the cipher image. The greater these criteria become, the better the JIEC scheme will operate. Equations 3 and 4 introduce these two criteria.

$$UACI = \frac{1}{N}\left[\sum_{j,k} \frac{C_1(j,k) - C_2(j,k)}{255}\right] \times 100\% \quad (3)$$

$$NPCR = \left(\frac{\sum_{j,k} D(j,k)}{N}\right) \times 100\% \quad (4)$$

In Equation 3, $C_1(j,k)$ and $C_2(j,k)$ show the values of pixels located in (j, k) on the two output images in comparison. N refers to the total number of pixels. The value of $D(j,k)$ can be determined through Equation 5 [14, 19, 21].

$$D(j,k) = \begin{bmatrix} 1 & C_1(i,j) \neq C_2(i,j) \\ 0 & otherwise \end{bmatrix} \quad (5)$$

In some papers, the appropriate magnitudes of NPCR and UACI were reported 99.61% and 33.46%, respectively [23].

*B. Efficiency Criteria*

These criteria include compression rate, signal-disturbance ratio, and the total runtime of a JIEC scheme, which are briefly discussed in this section.

The compression performance of a scheme is determined through CR (the ratio of the number of bits in the original image to the number of bits in the compressed image) and bpp (the ratio of total bits of the compressed image to the total pixels). A larger CR and a smaller bpp refer to a better compression [24, 25].

PSNR is employed to evaluate the quality of the recovered image (the effect of disturbance on the output algorithm). Equation 6 shows MSE, a criterion for determining the difference between the original image and the compressed one. According to the results[14], the higher the output quality of the scheme, the greater the PSNR. According to other results [19], if the PSNR of reconstructed images is greater than 30, the image quality is acceptable. Equations 6 and 7 can be employed to determine PSNR and MSE. In these equations, $X_{peak}$ shows the signal peak when $f(i,j)$ and $f^\wedge(i,j)$ refer to the number of pixels on the plain and cipher images, respectively, on the position indicated by (i, j). Finally, $M \times N$ shows the number of pixels [14, 24, 26].

$$MSE = \frac{1}{M \times N} \sum_{i=1}^{M} \sum_{j=1}^{N} (f^\wedge(i,j) - f(i,j))^2 \quad (6)$$

$$PSNR = 10 \times \log_{10}\left(\frac{M \times N \times 255^2}{MSE}\right) \quad (7)$$

Another criterion is the total runtime of the algorithm, which is very important in wireless communications. It means the encryption time plus the compression time spent on the entire algorithm.

IV. COMPARISON OF JIEC SCHEMES

The first goal of this comparison is to help users to select the best scheme meeting their requirements. Regarding the development of necessary and useful research context, the second goal is to introduce appropriate strategies for designing and developing novel JIEC schemes. In all of the schemes analyzed in this section, the comparison was drawn on the Lenna gray image (512 × 512).

Table 2 shows the general characteristics of the schemes evaluated in this section. Tables 3 and 4 discuss security and efficiency criteria, respectively.

According to Table 2, the key space ranges between $2^{176}$ and $2^{1024}$ on the compared schemes. If the space is too small, the scheme will be vulnerable against different attacks. However, if the key space becomes greater, the power consumption decreases in the total runtime.

In the scheme proposed by [16], the key size and the runtime are $2^{960}$ and 29 seconds, respectively, and the encryption time is 10.53 seconds regardless of the compression time. Therefore, not only was the encryption time appropriate, but the compression had an inappropriate performance lasting for nearly 18.47 seconds. However, the goal of compression is to reduce the image size and processing time.

In the scheme proposed by [8], the key space is $2^{554}$, and the encryption and compression time is 27 seconds altogether. The encryption lasted for nearly 9 seconds, whereas the compression lasted for 18 seconds, something which shows the inappropriate compression and encryption performances. In both schemes [8, 16], SPIHT and IWT structures were used for compression. According to the results, these two structures increased the compression time; thus, they are not recommended. Since compression and encryption are performed simultaneously in JIEC schemes, these two structures increase the total runtime.

In the scheme proposed by [11], the space key is $2^{464}$, and the total compression and encryption time is nearly 50 milliseconds. This runtime is acceptable.

In the scheme proposed by [15], the space key is $2^{176}$, and the scheme runtime is 4 seconds. Although it is shorter than those of the schemes proposed by [8, 16], it causes a long delay for wireless and online communications. The encryption time is nearly 3 milliseconds, considered appropriate.

In the scheme proposed by [12], the space key is nearly $2^{276}$, and it is nearly $2^{1024}$ in the scheme proposed by [13]. The scheme runtime is 28 milliseconds in DCT, which is shorter than those of other schemes. The necessary compression runtime is nearly 1 seconds for JPEG images [18]. In this scheme, the JIEC encryption and compression lasted for nearly 28 milliseconds in the DCT mode, whereas it lasted for nearly 56 milliseconds in the DWT mode. These results show the superiority of these schemes, in which compression was performed through compressive sensing when DCT, DWT, and DFT coefficients were used along with CA for encryption. According to the results, the use of compressive sensing and CRT took less runtime than the use of SPIHT and IWT.

According to Table 3, the schemes proposed by [8, 11, 16] showed higher values of entropy than those of other schemes due to using chaotic maps. In [8], different chaotic maps such as Lorenz, logistic, and Hénon maps were used for diffusion and confusion. In [11], a hyper-chaotic system was employed for diffusion and confusion. In [16], diffusion and confusion were based on the hyper-chaotic and cat map for encryption. Although all three schemes proposed by [13] showed acceptable and ideal levels of entropy, the results indicated that Lorenz, logistic, Hénon, cat, and Rabinovich encryption techniques caused higher levels of entropy than CA.

According to Table 3, the scheme proposed by [15] showed a higher NPCR, and the scheme proposed by [13] indicated a higher UACI than other papers. These results show that these schemes were more sensitive to pixel changes in the cipher image. The results also indicated that NPCR and UACI were 99.63 and 33.55, respectively, for the DFT mode in the scheme proposed by [13] in addition to the one proposed by [15]. All in all, these schemes were appropriate and resistant to differential attacks.

A JIEC scheme should be very key sensitive. Table 3 draws a comparison between key sensitivities of JIEC schemes. The scheme proposed by [13] showed the highest level of key sensitivity in DCT, DWT, and DFT modes when CA were used. The highest level (99.65%) was obtained in the DFT mode. It was an appropriate result. The key sensitivity was an appropriate level of 99.61% in the scheme proposed by [11], which employed the nonlinear hyper-chaotic dynamics and CRT to scramble pixels.

An image characteristic is the correlation between pixels, especially between the adjacent pixels. The JIEC scheme should be able to eliminate the correlation of images or make it approach null. According to the comparison results presented on Table 3, the scheme proposed by [13] showed the lowest level of correlation between the pixels of cipher images because it employed DFT-based CA to scramble pixels.

Regarding the comparison of JIEC schemes in CR and PSNR, compressive sensing compressors need less runtime than SPIHT and IWT compressors. In the schemes proposed by [15, 16], two different compressive sensing methods were used. For this purpose, three different classes of PSNR were considered:

$$PSNR \leq 30$$
$$30 < PSNR < 40$$
$$40 \leq PSNR$$

In the first class, KCS-DCT was the best scheme because of having a high PSNR and a more appropriate CR than those of others. In the second class, KCS-DFT was the best because of having more appropriate PSNR and CR than those of others. Finally, KCS-DCT was the best scheme in the third class. Accordingly, it appears that the scheme proposed by [13] was the best choice because it presented the lowest level of computation complexity due to using CR. It also showed high levels of confidentiality and flexibility. At the same time, it makes pixels too uniform. The next best scheme was the one proposed by [12]. It was resistant to disturbance and partial image loss.

V. SUGGESTIONS FOR FUTURE RESEARCH

According to the schemes analyzed in Section 4, a few suggestions and strategies are provided in this section to help researchers develop novel JIEC schemes meeting their different requirements. These suggestions and strategies are classified as the dos and don'ts of designing:

A. Dos of Designing

- Using certain structures such as DCT-based compressing sensing and the CRT simultaneously can increase compression speed and quality and also enhance the quality of the recovered image (Table 4).
- Regarding the encryption of JIEC schemes, the design speed can be enhanced by using Chen hyper-chaotic maps [12], the Chirikov encryption map [15], and CA based on DCT and DWT (Table 3).
- It is recommended to use CA, hyper-chaotic discrete nonlinear dynamics, and CRT (Table 3).
- It is recommended to employ the Chirikov map and CA to increase NPCR, UACI, and proper sensitivity to the plain image (Table 3).
- It is recommended to use DCT-based CA to decrease the correlation between adjacent pixels in the output image (Table 3).
- In order to increase entropy, it is recommended to use chaotic maps such as Lorenz, Logistic, Henon, Cat, and Rabinovich instead of CA (Table 3).

**Table 2.** Properties of Evaluated Schemes

| Reference | Encryption Method | Compression Method | Encryption (symmetric) | | Compression Type | | |
|---|---|---|---|---|---|---|---|
| | | | Stream cipher | Block | CS | Lossless | Lossy |
| [8] | OFB based on Lorenz, Hénon, and logistic maps | IWT, SPIHT | * | | | * | * |
| [11] | Permutation based on the 2D hyper-chaotic discrete nonlinear dynamic system | CRT | | * | | * | |
| [12] | Rotational shift and Chen map | 2D-CS | | * | * | | |
| [13] | Basic CA | KCS-DCT KCS-DWT KCS-DFT | | * | * | | |
| [14] | Cat map | DCT & orthogonal projection matrix | | * | | | * |
| [15] | Diffusion and confusion based on Chirikov standard map | DWT, SPIHT | | * | | * | * |
| [16] | Diffusion and confusion based on Rabinovich hyper-chaotic map | IWT, SPIHT | | * | | * | |
| [17] | Logistic map and Orthogonal Matrix | DCT | | * | | | * |
| [18] | Chen map knight's tour algorithm Semi Ham algorithm | DCT | * | | | | * |

**Table 3.** Comparison of Security Criteria

| References | | Key Space | Entropy | Total Runtime (s) | Key Sensitivity | NPCR | UACI | Horizontal Correlation | Orthogonal Correlation | Diagonal Correlation |
|---|---|---|---|---|---|---|---|---|---|---|
| [8] | | $2^{554}$ | 7.998 | 27 | 49.99 | - | - | 0.0116133 | 0.0063547 | 0.0018114 |
| [11] | | $2^{464}$ | 7.997 | 0.5 | 99.61 | - | - | 0.0058 | 0.0094 | 0.0214 |
| [12] | | $2^{276}$ | - | - | - | - | - | 0.0042 | -0.0043 | 0.0163 |
| [13] | DCT | $2^{1024}$ | 7.9219 | 0.2843 | 99.63 | 99.63 | 33.54 | 0.0026 | 0.0003 | 0.0012 |
| | DWT | $2^{1024}$ | 7.8986 | 0.561 | 99.63 | 99.63 | 33.51 | 0.0037 | 0.0018 | 0.0011 |
| | DFT | $2^{1024}$ | 7.9407 | 2.009 | 99.65 | 99.65 | 33.55 | 0.0024 | 0.0021 | 0.0009 |
| [14] | | $2^{186}$ | 7.998 | 25 | 99.61 | 99.40 | 0.3332 | 0.0132531 | 0.0010002 | 0.014118 |
| [15] | | $2^{176}$ | - | 3.56 | - | 99.91 | 33.51 | - | - | - |
| [16] | | $2^{960}$ | 7.992 | 29 | 49.6 | 99.59 | 33.43 | -0.0009732 | -0.0048559 | 0.0003075 |
| [17] | | $2^{225}$ | - | 0.9 | | 99.10 | 0.1538 | 0.9407 | -0.0140 | -0.0273 |
| [18] | | $2^{126}$ | - | 1.2068 | 99.51 | 99.61 | 33.465 | - | - | - |

**Table 4.** Comparison of Efficiency Criteria

| Reference | | | Lenna 256×256 | | Lenna 512×512 | | Baboon 512×512 | |
|---|---|---|---|---|---|---|---|---|
| | | | PSNR (Decibel) | CR (%) | PSNR (Decibel) | CR (%) | PSNR (Decibel) | CR (%) |
| [7] | | | - | - | - | 64 | - | 87 |
| [11] | | | - | - | - | 31 | - | 25 |
| [12] | | | 31 | 76 | - | - | - | - |
| | | | 25 | 56 | - | - | - | - |
| | | | 17 | 25 | - | - | - | - |
| [11] | DCT | | 45 | 72 | 45 | - | - | - |
| | | | 31 | 50 | 32 | - | - | - |
| | | | 27 | 25 | 28 | - | - | - |
| | DWT | | 35 | 75 | 34 | - | - | - |
| | | | 30 | 50 | 31 | - | - | - |
| | | | 26 | 25 | 27 | - | - | - |
| | DFT | | 25 | 25 | 25 | - | - | - |
| | | | 33 | 50 | 33 | - | - | - |
| | | | 44 | 75 | 44 | - | - | - |
| [14] | | | - | - | - | - | 31 | 75 |
| [15] | | | - | - | 77 | - | - | - |
| [16] | | | - | - | 77 | - | - | - |
| [17] | | | 32 | - | - | - | - | - |
| [18] | | | 10 | - | - | - | - | - |

Comparative study of Joint Image Encryption and Compression Schemes: A Review*B. Don'ts of Designing*

- It is not recommended to use certain methods such as SPIHT and IWT in online JIEC schemes due to their low compression speed and quality (Table 4).
- It is not recommended to employed Rabinovich, Hénon, Lorenz, and logistic chaotic maps in JIEC schemes due to their low speed of confusion and diffusion (Table 3).
- It is not recommended to use DFT in JIEC schemes due to its low compression speed (Table 4).
- Since Logistic map and orthogonal matrix create correlation in the output image pixels, and decrease NPCR and UACI values, it is not recommended to use them in the JIEC schemes (Table 3).

## VI. CONCLUSION

This paper contains a comparative study of the latest JIEC schemes. It considers a number of the most important criteria for the efficiency and security evaluation of JIEC schemes. Then some new schemes are compared together based on these criteria. Finally the comparison results were analysed to propose suggestions and strategies for future research to develop secure and efficient JIEC schemes. Also for future applications, it suggests that in order to choose an appropriate JIEC scheme, it is better to divide the applications and images into specific categories based on their attributes, and then select a proper scheme for each category.